\documentclass[journal,10pt]{IEEEtran}

\usepackage[dvips]{graphicx}
\usepackage[latin1]{inputenc}
\usepackage{url}
\usepackage{multicol}
\usepackage{cite}
\usepackage{stfloats}
\usepackage{amsmath,amsfonts,amssymb}
\usepackage{algorithm}
\usepackage{algpseudocode}

\newcommand{\bb}{\mathbf{b}}

\newcommand{\thB}{\mbox{\boldmath$\theta$}}
\newcommand{\thBs}{\mbox{\tiny $\theta$}}

\newcommand{\Ms}{\mbox{\tiny $M$}}

\begin{document}
\title{Cramer-Rao bound for source estimation using a network of binary sensors}
\author{Branko~Ristic\thanks{Authors are with DSTO, LD, Bld. 94, Melbourne, VIC 3207,
Australia. The corresponding author is B. Ristic, email:
branko.ristic@dsto.defence.gov.au; tel: +61 3 9626 8370;
fax: +61 3 9626 8473. }, 
        Ajith~Gunatilaka, Ralph Gailis
}

\maketitle

\begin{abstract}
The paper derives the theoretical Cramer-Rao lower bound for
parameter estimation of a source (of emitting energy,  gas,
aerosol), monitored by a network of sensors providing binary
measurements. The theoretical bound is studied in the context of a
source of a continuous release in the atmosphere of hazardous gas or
aerosol. Numerical results show a good agreement with the empirical
errors, obtained using an MCMC parameter estimation technique.
\end{abstract}

\begin{IEEEkeywords}
Binary sensor network, Cramer-Rao lower bound, source localisation,
dispersion model
\end{IEEEkeywords}

\section{Introduction}

Binary sensor networks have become widespread in environmental
monitoring applications because binary sensors generate as little as
one bit of information, thereby providing inexpensive sensing with
minimal communication requirements \cite{aslam_03}. The motivation
for our study is the theoretical prediction of the best achievable
accuracy in localisation of a source of hazardous release of gas or
aerosols, using such a binary sensor network. However, the
formulation of the problem will be general enough to be applicable
to parameter estimation of any emitting source, including the source
of sound, vibration, seismic activity, radiation, etc.

The paper derives the theoretically smallest achievable second-order
estimation error in the form of the Cram\'{e}r-Rao lower bound
\cite{vantrees_68}. The derivation is carried out in the Bayesian
framework, that is, assuming that some prior knowledge of source
parameters is available. To our best knowledge, this type of
Cram\'{e}r-Rao bound (CRB), for source parameter estimation using
binary sensors, has not been derived earlier. The closest references
are \cite{niu_06,ribeiro_06,djuric_08}. In \cite{niu_06}, a CRB is
derived for a quantised sensor network in the context of target
localisation. However, the bound in \cite{niu_06} is limited to the
received signal strength (RSS) measurement model only. Hence, the
bound we derive is more general, albeit restricted to binary
quantisation.  A special case of the CRB we derive appeared in
\cite{ribeiro_06}. Finally, a CRB for tracking a moving target using
a binary sensor network and RSS  measurements was presented in
\cite{djuric_08}, although it is not clear how and where the
likelihood function of binary sensors was used in derivation.

\section{Problem statement}
The problem is to derive the lower bound of estimation error for the
parameter vector  $\thB\in\mathbb{R}^M$. In the context of source
estimation, the parameter vector typically includes not only the
source parameters, such as its location (coordinates), size, and the
release-rate (intensity), but also the propagation and measurement
model parameters, such as the attenuation factors, meteorological
parameters, and sensor characteristics. The measurement at $i$th
sensor, $i=1,2,\dots,S$, is a scalar (e.g. concentration of the gas,
the amount of received energy). Before it is binary quantised, the
``analog'' measurement is modelled by:
\begin{equation}
z_i = C_i(\thB)+w_i  \label{e:mod1}
\end{equation}
where \begin{itemize} \item $C_i(\thB)$ is the dispersion or
propagation measurement model, which includes the sensor index $i$
in the subscript being a function of the sensor location;
\item $w_i$ is additive white zero-mean Gaussian noise, independent
of noise processes in other sensors: $w_i \sim f(w) =
1/(2\pi\sigma)\exp[-w^2/(2\sigma^2)]$.
\end{itemize}
The actual measurement supplied by sensor $i$ is binary, that is:
\begin{equation}
b_i = \begin{cases} 1 & \text{ if } z_i > \tau \\
                    0 & \text{ if } z_i \leq \tau,
                    \end{cases}
                    \end{equation}
                    where $\tau$ is the threshold.
The probability of binary measurement $b_i=1$ can then be expressed
as:
\begin{equation}
q_i(\thB) = Pr\{b_i = 1|\thB\}  =  F(\tau - C_i(\thB))
\end{equation}
where $F(x) = 1/(\sqrt{2\pi}\sigma)\int_{x}^{\infty}
e^{-\frac{u^2}{2\sigma^2}} du$ is the complementary cumulative
distribution function of Gaussian noise.

Let us now group all binary sensor measurements into a vector: $\bb
= [b_1,\cdots,b_S]^\intercal$. The likelihood function for the
binary measurement vector is then:
\begin{equation}
p(\bb|\thB) = \prod_{i=1}^S [q_i(\thB)]^{b_i}\,
[1-q_i(\thB)]^{1-b_i}. \label{e:like}
\end{equation}

Assuming the prior probability density function (pdf) of the
parameter vector is known and denoted $\pi(\thB)$, the objective of
Bayesian estimation is to determine the posterior density
\begin{equation}
p(\thB|\bb) \propto p(\bb|\thB)\, \pi(\thB). \label{e:bayes1}
\end{equation}
Bayesian estimators of $\thB$ (e.g. the expected a posteriori or the
maximum a posteriori) can then be computed from the posterior
$p(\thB|\bb)$.

The Cram\'{e}r-Rao lower bound states that the covariance matrix of
an unbiased estimator $\hat{\thB}$  of the parameter vector is
bounded from below as follows \cite{vantrees_68}:
\begin{equation}
\mbox{E}\left\{\left(\hat{\thB} - \thB^*\right)\;\left(\hat{\thB} -
\thB^*\right )^\intercal \right \} \geq {\bf
  J}^{-1},
\label{e:fim1}
\end{equation}
where $\thB^*$ is the true value of the parameter vector and ${\bf
J}$ is the information matrix, defined as
\begin{equation}
\mathbf{J} = - \mathbb{E}\left\{\nabla_{\thBs}
\nabla_{\thBs}^\intercal \log p(\thB|\bb)\right\}. \label{e:IMdef}
\end{equation}
Operator $\nabla_{\thBs}$, which  features in (\ref{e:IMdef}), is
the gradient with respect to $\thB$: if we denote the $n$th
component of vector $\thB$ by $\theta_n$, keeping in mind that
$\text{dim}(\thB) = M$, then
\begin{equation}
\nabla_{\thBs} = \left[\partial / \partial \theta_1, \;\; \cdots,
\partial / \partial \theta_{\Ms} \right]^\intercal.
\end{equation}
Expression (\ref{e:IMdef}) is evaluated at the true value of the
parameter vector $\thB^*$. The expectation operator $\mathbb{E}$ in
(\ref{e:IMdef}) is w.r.t. the binary measurement vector $\bb$.

Our goal is to derive the analytic expression for the information
matrix ${\bf J}$ as a function of $C_i(\thB)$, $f(x)$, $F(x)$, and
$\tau$. Then the CRB will follow as the inverse matrix of ${\bf J}$.

\section{Derivation of the information matrix}

Substitution of (\ref{e:bayes1}) into (\ref{e:IMdef}) leads to:
\begin{equation}
\mathbf{J} = \underbrace{-\mathbb{E}\left\{\nabla_{\thBs}
\nabla_{\thBs}^\intercal \log p(\bb|\thB)\right\}}_{\mathbf{J}^d}
 \underbrace{-\mathbb{E}\left\{\nabla_{\thBs}\nabla_{\thBs}^\intercal
\log \pi(\thB)\right\}}_{\mathbf{J}^p} \label{e:pd}
\end{equation}
where $\mathbf{J}^d$ and $\mathbf{J}^p$ are the information matrices
corresponding to the measurements (data) and the prior,
respectively. If we adopt for convenience a Gaussian prior, i.e. $
\pi(\thB) = \mathcal{N}(\thB; \thB^*,\mathbf{\Sigma})$, with a
diagonal covariance matrix $\mathbf{\Sigma}$, then $\mathbf{J}^p =
\mathbf{\Sigma}^{-1}$. The CRB is according to (\ref{e:fim1}) and
(\ref{e:pd}) defined as  $(\mathbf{J}^d + \mathbf{J}^p)^{-1}$, and
is often referred to as the {\em posterior} CRB, in order to
emphasize that it includes the contributions from both the prior and
the measurements.

In order to derive the expression for $\mathbf{J}^d =
-\mathbb{E}\left\{\nabla_{\thBs} \nabla_{\thBs}^\intercal \log
p(\bb|\thB)\right\}$, note first that $\nabla_{\thBs}
\nabla_{\thBs}^\intercal \equiv \triangle_{\thBs}$ is the Hessian
operator with respect to $\thB$:
\begin{equation}
\triangle_{\thBs} = \left[\begin{matrix} \frac{\partial^2 }{\partial
\theta_1^2} & \cdots &
\frac{\partial^2 }{\partial \theta_1\partial \theta_{\Ms}}  \\
\vdots & \ddots  & \vdots \\
\frac{\partial^2 }{\partial \theta_{\Ms}\partial \theta_1} & \cdots
& \frac{\partial^2 }{\partial \theta_{\Ms}^2} \end{matrix} \right]
\label{e:hess}
\end{equation}
Next, let us write the expression for the log-likelihood function,
which follows from (\ref{e:like}):
\begin{equation}
\log p(\bb|\thB) = \sum_{i=1}^S \big[b_i\log q_i(\thB) + (1-b_i)
\log (1-q_i(\thB)) \big]
\end{equation}
After a few steps of mathematical manipulations it can be shown that
the first partial derivative of the log-likelihood is:
\begin{align}
\frac{\partial \log p(\bb|\thB)}{\partial \theta_m}= &
    \sum_{i=1}^S\bigg[-b_i\,
    \frac{f(\tau-C_i(\thB))}{F(\tau-C_i(\thB))}\,
    \frac{\partial C_i(\thB)}{\partial \theta_m} \nonumber \\ & +
    (1-b_i)\frac{f(\tau-C_i(\thB))}{1-F(\tau-C_i(\thB))}\, \frac{\partial C_i(\thB)}{\partial \theta_m}\bigg]
\end{align}
for $m=1,\dots,M$. Likewise, the second partial derivatives, which
feature in Hessian (\ref{e:hess}), are given by:
\begin{align}
\frac{\partial^2 \log p(\bb|\thB)}{\partial \theta_m
\partial \theta_n}  & =
    \sum_{i=1}^S\bigg\{ b_i \bigg[- \frac{\mathcal{A}}{F^2(\tau-C_i(\thB))} \nonumber \\
    & + \frac{\mathcal{B}}{F(\tau-C_i(\thB))} -
    \frac{\mathcal{C}}{F(\tau-C_i(\thB))} \bigg] \nonumber \\
    & + (1-b_i)\bigg[ -\frac{\mathcal{A}}{(1-F(\tau-C_i(\thB)))^2} \nonumber \\ &- \frac{\mathcal{B}}{1-F(\tau-C_i(\thB))} +
    \frac{\mathcal{C}}{1-F(\tau-C_i(\thB))}\bigg] \bigg\}
\end{align}
for any $m,n=1,2,\dots,M$, with:
\begin{eqnarray}
\mathcal{A} & = & f^2(\tau-C_i(\thB))\,\frac{\partial
C_i(\thB)}{\partial \theta_m} \frac{\partial C_i(\thB)} {\partial
\theta_n} \nonumber \\
\mathcal{B} & = & f'(\tau-C_i(\thB))\,\frac{\partial
C_i(\thB)}{\partial \theta_m}\,
    \frac{\partial C_i(\thB)}{\partial \theta_n} \nonumber \\
\mathcal{C} & = & f(\tau-C_i(\thB))\,\frac{\partial^2
C_i(\thB)}{\partial \theta_m \partial \theta_n}.
\end{eqnarray}

 After taking the expectation over $b_i$, using the fact that
$\mathbb{E}[b_i]=q_i(\thB)=F(\tau-C_i(\thB))$, followed by
simplification, we obtain for the $(m,n)$th element of matrix
$\mathbf{J}^d$:
\begin{eqnarray}
J^d_{m,n} & = &  - \mathbb{E}\left\{\frac{\partial^2 \log
p(\bb|\thB)}{\partial \theta_m
\partial \theta_n}\right\} \nonumber \\
& = & \sum_{i=1}^S
\frac{f^2(\tau-C_i(\thB))}{q_i(\thB)(1-q_i(\thB))} \;\frac{\partial
C_i(\thB)}{\partial \theta_m}\; \frac{\partial C_i(\thB)}{\partial
\theta_n} \label{e:final}
\end{eqnarray}

A special case of the information matrix $\mathbf{J}^d$, for $M=1$
and $C_i(\thB) = \theta$, was derived in \cite{ribeiro_06}. Since in
this case all $q_i(\theta)$, $i=1,\dots,S$ are equal and denoted
$q(\theta)$, from (\ref{e:final}) it follows that:
\[ J^d = S \frac{f^2(\tau-\theta))}{q(\theta)(1-q(\theta))}. \]
This expression appears in eq.(7) of  \cite{ribeiro_06}.

Another special case is the source localisation using binary RSS
measurements, where the measurement model is
\cite{patwari_05,niu_06}: $C_i(\thB) = Q_0 - 20\log
(\sqrt{(x_i-x_0)^2+(y_i-y_0)^2}/d_0)$. The parameter vector $\thB =
[x_0, y_0, Q_0]^\intercal$ includes the source coordinates
$(x_0,y_0)$ and its intensity $Q_0$. The coordinates of the $i$th
sensor are $(x_i,y_i)$. The CRB for this case has been derived in
\cite{niu_06}.

\begin{figure}[htb]
\centerline{\includegraphics[height=5.6cm]{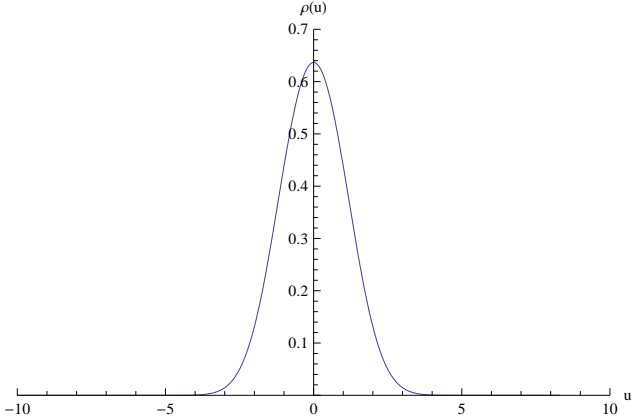}}
 \caption{Ratio $\rho(u)$ for $\sigma=1$ over $-10\leq u \leq 10$, see (\ref{e:rho}). }
\label{f:rho}
\end{figure}

For completeness, we point out that if the analog (non-quantised)
measurements $z_i$, $i=1,\dots,S$ of (\ref{e:mod1}) are used for
source estimation, the expression for the $(m,n)$th element of the
information matrix (due to data) is given by \cite{ssp_crb_14}:
\begin{equation}
\tilde{J}^d_{m,n} =\frac{1}{\sigma^2}\sum_{i=1}^S \frac{\partial
C_i(\thB)}{\partial \theta_m}\; \frac{\partial C_i(\thB)}{\partial
\theta_n} \label{e:final0}
\end{equation}
Comparing (\ref{e:final}) to (\ref{e:final0}) one can note that the
only difference is that the ratio
\begin{equation}
\rho(\tau-C_i(\thB)) =
\frac{f^2(\tau-C_i(\thB))}{F(\tau-C_i(\thB))[1-F(\tau-C_i(\thB))]},
\label{e:rho}
\end{equation}
which features in (\ref{e:final}), is replaced by $1/\sigma^2$ in
(\ref{e:final0}). Using the L'Hopital's rule, it can be shown that
$\lim_{u\rightarrow \pm \infty} \rho(u) = 0$. The implication is
that, if the threshold $\tau$ is too high or too low, the binary
measurements become uninformative.  Fig.\ref{f:rho} displays a plot
of $\rho(u)$ for $\sigma=1$. Observe that $\rho(u)$ reaches its
maximum at $u=0$; this maximum, however, is smaller than the factor
$1/\sigma^2 = 1$, which according to (\ref{e:final0}) appears in the
analog signal case. The conclusion is that the CRB for a binary
sensor network, is always higher (irrespective of the threshold)
than the CRB for the corresponding analog sensor network.

Next we consider a practical application of the CRB for binary
sensor networks.


\section{Application: Biochemical source localisation}

\subsection{The measurement model and its derivatives}

Localisation of a source of hazardous biochemical material, released
in the atmosphere and transported by wind, is very important for
public safety \cite{zhao_nehorai_06}. The measurement model
$C_i(\thB)$ in this application is a suitable atmospheric dispersion
model \cite{Arya1998}.  Such a model describes via mathematical
equations the physical processes that govern the atmospheric
dispersion of biological pathogens or chemical substances within the
plume. We adopt  in this study the Gaussian plume model, being the
core of all regulatory atmospheric dispersion models
\cite{Arya1998}.

Suppose a biochemical source  is located at coordinates
$(x_0,y_0,z_0)$. The release rate of the source is $Q_0$. By
convention, the wind direction coincides with the direction of the
$x$ axis. The mean wind speed is denoted by $U$; the spread of the
plume in $y$ and $z$ direction for $x>x_0$ is modelled by
\cite{venkatram_05}
\begin{eqnarray}
\sigma_y & = &  \sigma_v (x-x_0)/U \\
\sigma_z & = &  \sigma_w (x-x_0)/U,
\end{eqnarray}
respectively, where  $\sigma_v$ and $\sigma_w$ are environmental
parameters. In reality,  $x_0$, $y_0$,  $z_0$, $Q_0$, $U$,
 $\sigma_v$ and $\sigma_w$ are unknown parameters,
although  prior knowledge is available for some of them in the form
of meteorological/environmental advice. For simplicity, however, we
will focus on localisation only, that is, only the source
coordinates $x_0$ and $y_0$ are assumed unknown, hence $\thB
=\left[x_0\; y_0\right]^\intercal$. The Gaussian plume model of a
concentration measurement at $i$th sensor, $i=1,\dots,S$, located at
coordinates $(x_i>x_0,y_i,z_i=0)$, is given by \cite{venkatram_05}
\begin{equation}
C_i(\thB) = \frac{Q_0}{\pi\sigma_{y_i}\sigma_{z_i}
U}\exp\left\{-\frac{z_0^2}{2\sigma_{z_i}^2}\right\}\exp\left\{-\frac{(y_i-y_0)^2}{2\sigma_{y_i}^2}\right\}.
\label{e:disper}
\end{equation}
Note that the plume spreads $\sigma_{y_i}$ and $\sigma_{z_i}$ in
(\ref{e:disper}) are assigned the sensor index $i$, because they are
computed at $x_i$.

In order to compute the information matrix (and the CRB), according
to (\ref{e:final}), we need to derive the partial derivatives
$\partial C_i(\thB)/\partial x_0$ and $\partial C_i(\thB)/\partial
y_0$. After few steps one can get \cite{ssp_crb_14}:
\begin{equation}
\frac{\partial C_{i}(\thB) }{\partial x_0} = \alpha + \beta+\gamma
\end{equation}
with
\begin{align*}
\alpha &= \frac{Q_0\sigma_w
e^{-\frac{z_0^2}{2\sigma_{z_i}^2}}e^{-\frac{(y_i-y_0)^2}{2\sigma_{y_i}^2}}}{\pi
U^2\sigma_{y_i} \sigma_{z_i}^2},\hspace{1mm} \beta  =
\frac{Q_0\sigma_v
e^{-\frac{z_0^2}{2\sigma_{z_i}^2}}e^{-\frac{(y_i-y_0)^2}{2\sigma_{y_i}^2}}}{\pi
U^2\sigma_{y_i}^2 \sigma_{z_i}}\\
\gamma & =  \frac{Q_0
e^{-\frac{z_0^2}{2\sigma_{z_i}^2}}e^{-\frac{(y_i-y_0)^2}{2\sigma_{y_i}^2}}}{\pi
U\sigma_{y_i} \sigma_{z_i}}  \left[
-\frac{(y_i-y_0)^2\sigma_v}{U\sigma_{y_i}^3}
-\frac{z_0^2\sigma_w}{U\sigma_{z_i}^3}\right]
\end{align*}
Similarly,
\begin{equation}
\frac{\partial C_{i}(\thB) }{\partial y_0} = \frac{Q_0\,(y_i-y_0)\,
e^{-\frac{z_0^2}{2\sigma_{z_i}^2}}e^{-\frac{(y_i-y_0)^2}{2\sigma_{y_i}^2}}}{\pi
U\sigma_{y_i}^3 \sigma_{z_i}}
\end{equation}

\subsection{Numerical analysis and verification}

\begin{figure}[htb]
\centerline{\includegraphics[height=5.6cm]{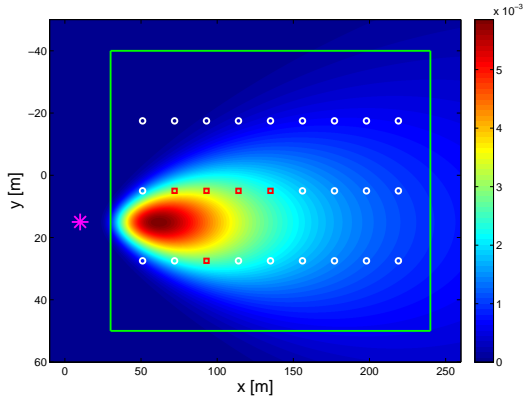}}
 \caption{The scenario with $S=27$ sensors for numerical analysis and verification of the CRB. }
\label{f:1}
\end{figure}

The CRB is computed and verified for a scenario plotted in
Fig.\ref{f:1}. The source is marked by the asterisk at
 coordinates $(10,15)$. The colours indicate the level of concentration of the released material on the ground, i.e. $z_i=0$.
 The area populated by binary sensors is indicated by a rectangle
 whose lower-left corner is at $(30,-40)$ m, and the upper right corner
 at $(240,50)$ m.  The total number of binary sensors is $S=27$.
 The locations of sensors with measurements $b_i=1$ are marked by red squares, while
 those with $b_i=0$ are indicated by white circles, using threshold $\tau= 0.0024$. Other parameters used in
  the simulation are as follows: $z_0=5$ m, $Q_0= 5$ g/s,  $U=3.5$
  m/s,  $\sigma_v = 0.5$ m/s, $\sigma_w = 0.2$ m/s. The standard
  deviation of noise $\sigma =  0.0001$ g/m$^3$. The covariance
  matrix of the prior pdf is $\mathbf{\Sigma} = \mbox{diag}[500^2,\;
  500^2]$.

\begin{figure}[htb]
\centerline{\includegraphics[height=5.3cm]{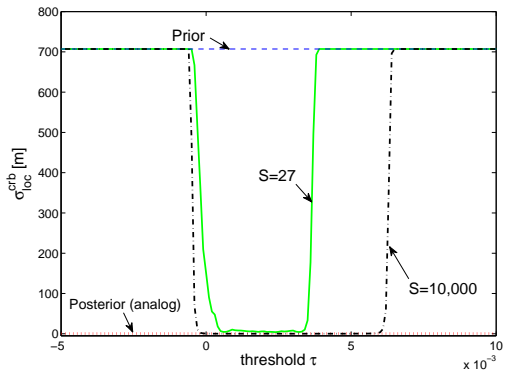}}
 \caption{Standard deviation of localisation error using a binary sensor network,
 computed from the theoretical posterior CRB and plotted as a function of threshold $\tau$:
 the solid green line is for $S=27$ sensors; the dash-dotted line is for $S=10,000$ sensors}
\label{f:2}
\end{figure}

The posterior CRB, $\mathbf{J}^{-1}$, in this case is a $2\times 2$
matrix, from which we can express the theoretically best achievable
standard deviation of localisation error as
\begin{equation}
\sigma^{\mbox{\tiny crb}}_{loc} =
\sqrt{\mbox{tr}\left[\mathbf{J}^{-1}\right]}.  \label{e:sig_b}
\end{equation}
This posterior standard deviation, as a function of the threshold
$\tau$,  is plotted by a solid green line in Fig.\ref{f:2} for the
adopted scenario with binary sensors. The horizontal blue dashed
line at $\sigma^{\mbox{\tiny
crb}}_{loc}=\sqrt{\mbox{tr}\left[\mathbf{J^p}^{-1}\right]} =
\sqrt{2\times 500^2}\approx 707$ m indicates the prior standard
deviation of source location error; the horizontal red dotted line
at $\sigma^{\mbox{\tiny crb}}_{loc}\approx 0.3$ m marks the value
computed using the CRB for analog (non-quantised) measurements, via
(\ref{e:final0}).

Note that, as discussed earlier, for too high and too low threshold
values $\tau$, the posterior localisation uncertainty equals the
prior uncertainty, because the information contained in the binary
measurements equals zero (this is when all measurements are either
zero or one). For a middle range of $\tau$ values, the posterior
standard deviation of binary measurements approaches the posterior
standard deviation of analog measurements (but never reaches it, as
discussed earlier). This observation applies even when the number of
sensors is increased, as demonstrated by the dash-dotted line in
Fig.\ref{f:2}: this line shows the posterior standard deviation of
binary measurements $\sigma^{\mbox{\tiny crb}}_{loc}$ for
$S=200\times 50=10,000$ sensors placed on a uniform grid inside the
rectangular area indicated in Fig.\ref{f:1}.

The theoretical bounds are next compared with the empirical
estimation errors obtained using a Markov chain Monte Carlo (MCMC)
based parameter estimation algorithm \cite{robert_casella}. The MCMC
algorithm is initialised by repeatedly drawing samples (candidate
source coordinates) from the prior $\pi(\thB)$ until $n_s$ samples,
whose likelihood (\ref{e:like}) is greater than zero, are found. The
sample with the highest value of the likelihood is selected as the
starting point of the Metropolis-Hastings algorithm. The proposal
distribution of the MCMC is Gaussian with the mean equal to the
current sample and the covariance matrix equal to the theoretical
CRB (details of the MCMC algorithms are omitted). The source
location estimate is computed as the mean value of the last $n_m$
samples generated by the MCMC. Our practical implementation used the
following values: $n_s=10$ and $n_m=10000$. Table \ref{t:1} shows
the results for the parameter values as listed above, using the
threshold  $\tau=0.0018$. Three different sensor placements are
considered:\\
$\bullet$ {\em Placement 1}:  $S=16$ sensors, with sensor
$x$-coordinate $x_i\in\{40,100,160,220\}$m and $y$-coordinate
$y_i\in\{-20,0,20,40\}$m; this placement is contained in placements 2 and 3.\\
$\bullet$ {\em Placement 2}: $S=28$ sensors, with $x_i\in\{40,$
$70,$ $100,$ $130,$ $160,$ $190,$ $220\}$m and $y_i\in\{-20,$
$40,$ $20,$ $40\}$m; this placement is contained in placement 3.\\
$\bullet$ {\em Placement 3}: $S=49$ sensors, with $x_i\in \{40,$
$70,$ $100,$ $ 130,$ $160,$ $190,$ $ 220\}$m and
$y_i\in\{-20,-10,0,10,20,30,40\}$m.

Table \ref{t:1} demonstrates a good agreement between the
theoretical value $\sigma^{\mbox{\tiny crb}}_{loc}$ of
(\ref{e:sig_b}) and the root-mean-squared (RMS) error
$\widehat{\epsilon}_{loc}$ resulting from the MCMC localisation. The
RMS error is computed as:
\begin{equation}
\widehat{\epsilon}_{loc} = \sqrt{\frac{1}{L}\sum_{\ell=1}^L
\left[(\hat{x}^\ell_0 - x_0)^2 + (\hat{y}^\ell_0 - y_0)^2 \right]}
\end{equation}
where $(\hat{x}^\ell_0,\hat{y}^\ell_0)$ are MCMC estimated
coordinates of the source at the $\ell$th Monte Carlo run, with
$\ell=1,\dots,L$ and $L=200$ is the total number of  Monte Carlo
runs.

\begin{table}
 \caption{Theoretical CRBs compared with
 the root mean square errors of an MCMC algorithm, averaged over 200 Monte Carlo runs}
\begin{center}
\begin{tabular}{ccc}\hline\hline
Sensor  & Theoretical CRB  &  RMS error  \\
placement & $\sigma^{\mbox{\tiny crb}}_{loc}$ & $\widehat{\epsilon}_{loc}$   \\
 \hline\hline $(1)$ & $5.75$ m & $7.33$ m
\\
\hline $(2)$ & $3.93$ m &  $4.08$ m \\ \hline $(3)$  & $0.68$ m &
$2.55$ m
\\ \hline
 \hline
\end{tabular}
\end{center}
\label{t:1}
\end{table}

\section{Summary}

The paper derived the theoretical Cram\'{e}r-Rao lower bound for
source estimation using measurements collected by a binary sensor
network. The key result, given by (\ref{e:final}), appears
surprisingly simple and elegant. The bound is studied numerically in
the context of a source of biochemical tracer (aerosol, gas)
released in the atmosphere and transported by wind. Using a Gaussian
plume dispersion model, the paper computed the theoretical bound and
found that it approaches (but never reaches) the corresponding bound
for analog (non-quantised) measurements, if the binary threshold is
chosen properly. Finally, a good agreement between the theoretical
bound and empirical errors (obtained using an MCMC based parameter
estimation algorithm) is established.

\bibliographystyle{IEEEbib}

\end{document}